\documentclass[letterpaper, 10 pt, conference]{ieeeconf}  

\IEEEoverridecommandlockouts                              

\usepackage{graphicx}
\usepackage{tabularx}
\usepackage{url}
\usepackage{multirow}
\usepackage{longtable}
\usepackage{booktabs}
\usepackage{caption}

\title{\LARGE \bf
Towards Emotion Co-regulation with LLM-powered Socially Assistive Robots: Integrating LLM Prompts and Robotic Behaviors to Support Parent-Neurodivergent Child Dyads
}

\author{Jing Li$^{1}$ and Felix Schijve$^{2}$ and Sheng Li$^{3}$ and Yuye Yang$^{4}$ and Jun Hu$^{5}$ and Emilia Barakova$^{6}$
\thanks{$^{1}$Jing Li is with the Department of Industrial Design,
        Eindhoven University of Technology, 5612 AZ Eindhoven, The Netherlands
        {\tt\small j.li2@tue.nl}}%
\thanks{$^{2}$Felix Schijve is with the Department of Biomedical Engineering,
        Eindhoven University of Technology, 5612 AZ Eindhoven, The Netherlands
        {\tt\small felixschijve@hotmail.com}}%
\thanks{$^{3}$Sheng Li is with the Department of Engineering,
        Institute of Science Tokyo, Yokohama, Japan
        {\tt\small sheng.li@ieee.org}}%
\thanks{$^{4}$Yuye Yang is with the Department of Engineering,
        Utrecht University, 3584 CS Utrecht, The Netherlands
        {\tt\small y.yang14@students.uu.nl}}%
\thanks{$^{5}$Jun Hu is with the Department of Industrial Design,
        Eindhoven University of Technology, 5612 AZ Eindhoven, The Netherlands
        {\tt\small j.hu@tue.nl}}%
\thanks{$^{6}$Emilia Barakova is with the Department of Industrial Design,
        Eindhoven University of Technology, 5612 AZ Eindhoven, The Netherlands
        {\tt\small e.i.barakova@tue.nl}}%
}

\begin{document}

\maketitle
\thispagestyle{empty}
\pagestyle{empty}

\begin{abstract}

Socially Assistive Robotics (SAR) has shown promise in supporting emotion regulation for neurodivergent children. Recently, there has been increasing interest in leveraging advanced technologies to assist parents in co-regulating emotions with their children. However, limited research has explored the integration of large language models (LLMs) with SAR to facilitate emotion co-regulation between parents and children with neurodevelopmental disorders. To address this gap, we developed an LLM-powered social robot by deploying a speech communication module on the MiRo-E robotic platform. This supervised autonomous system integrates LLM prompts and robotic behaviors to deliver tailored interventions for both parents and neurodivergent children. Pilot tests were conducted with two parent-child dyads, followed by a qualitative analysis. The findings reveal MiRo-E's positive impacts on interaction dynamics and its potential to facilitate emotion regulation, along with identified design and technical challenges. Based on these insights, we provide design implications to advance the future development of LLM-powered SAR for mental health applications.

\end{abstract}

\section{INTRODUCTION}
\subsection{Emotion co-regulation in parent-neurodivergent child dyads}
Neurodivergent children often experience difficulties with emotion regulation (ER) \cite{thompson2007emotion,ting2017emotion}, which refers to the ability to manage emotions, attention, and stress. For these children with neurodevelopmental disorders such as  Autism Spectrum Disorder (ASD) and Attention Deficit Hyperactivity Disorder (ADHD), parental emotion co-regulation plays a pivotal role \cite{silva2023unpacking,gulsrud2010co}. This process involves parents supporting their child’s emotional development through motivational or emotional scaffolding and employing appropriate regulation strategies \cite{gulsrud2010co,ting2017emotion}. Research consistently highlights a strong positive correlation between the quality of parent-child emotion co-regulation and the development of ER skills in neurodivergent children \cite{ting2017emotion}. However, co-regulation in the context of neurodevelopmental disorders is often a demanding and stressful endeavor, requiring substantial emotional resilience and self-regulation skills from parents \cite{hayes2013impact,sanders2018importance}. To address these challenges, recent studies have emphasized the need for advanced technological solutions to support parents in providing emotion co-regulation for neurodivergent children  \cite{silva2024co,piccolo2024parental}. Various technological interventions have been explored, including mobile applications \cite{silva2023unpacking,sonne2016changing}, biofeedback agents \cite{li2024stress,wang2024evaluating}, and interactive toys \cite{iskanderani2023research,chan2017wakey}. Among these, socially assistive robotics (SAR) has demonstrated substantial potential to support neurodivergent children and to facilitate parent-child co-regulation in a wide range of contexts \cite{gvirsman2024effect,theofanopoulou2022exploring,ho2023designing}.

\subsection{Socially assistive robots in therapy} 
Socially assistive robots are designed to provide support through social interactions with users \cite{feil2005defining}. Their embodied presence offers a more engaging and effective alternative to screen-based interventions \cite{barakova2018socially,kabacinska2021socially}, by leveraging physical appearance, multiple interaction modalities \cite{feng2022context}, the ability to provide speech aligned with nonverbal behaviors \cite{van2021preferences}, and even matched personalities \cite{andriella2021have}. However, most SAR therapies aimed at supporting the mental health of neurodivergent children rely on remote-controlled operation using the Wizard of Oz (WoZ) paradigm. This method allows for a higher level of social interaction without requiring advanced artificial intelligence \cite{esteban2017build}, which is not sustainable due to its high cost and time-consuming nature \cite{scassellati2012robots}. Consequently, researchers have proposed that SAR used in therapeutic scenarios should be capable of operating autonomously \cite{thill2012robot}. Esteban et al. \cite{esteban2017build} introduced the supervised autonomy approach for SAR therapies, in which human operators (e.g. therapists, psychologists or teachers) define and provide specific goals for the robot, allowing it to function autonomously while working toward these goals.

\subsection{LLM-powered social robots for emotion support} 
Recent advances in Large Language Models (LLMs) have further expanded the potential of SAR by enabling more context-aware and versatile human-robot interactions. These models have revolutionized chatbot development, allowing for more natural, conversational exchanges that closely resemble human communication \cite{jo2023understanding}. Emerging studies indicate that LLMs can exhibit emotional intelligence comparable to humans \cite{wang2023emotional}, with their responses often perceived as balanced, empathetic, and supportive \cite{menon2023chatting}. Furthermore, LLMs have been investigated for applications such as social scripting and emotional support \cite{cha2021exploring,choi2024unlock}, showing their promise in enhancing the capabilities of SAR to address the nuanced emotional needs of neurodivergent families.

While LLMs excel in language comprehension and generation, they lack the ability to engage in real-time, embodied interactions within physical environments \cite{xi2023rise}. In contrast, SAR is designed to operate in physical contexts, relying on precise sensory-motor feedback and contextual reasoning to interact effectively \cite{nault2024socially}. Bridging the gap between these technologies requires the development of systems that integrate language-based reasoning with embodied actions \cite{shi2024can}. Such systems must ensure reliable, explainable, and context-aware robot control, enabling SAR to dynamically and meaningfully support emotion co-regulation in the complex context of parent-child interactions \cite{shi2024can,esteban2017build,kim2024understanding}. Despite the immense potential of LLM-powered robots to transform human-robot interaction, research remains limited in identifying the specific design requirements and implications of integrating LLMs into socially assistive robots for therapeutic applications.

\subsection{Research Goals and Contributions} 
Motivated by the research gap, our goal is to explore strategies for integrating LLM Prompts and robotic behaviors to support emotion co-regulation between parents and children with neurodevelopmental disorders. To explore the integration of LLMs and SAR for assisting emotion co-regulation in triadic interactions, we introduce the design of an LLM-powered social robot aimed at facilitating emotion co-regulation during stressful parent-child collaboration tasks, and evaluate this design through two pilot tests, addressing the following research questions: 1.) To what extent is the designed integration of LLM prompts and robotic behaviors effective in supporting triadic interactions? 2.) How do parent-child dyads engage with and adhere to the robotic interventions?

In this paper, we present the design of an LLM-powered social robot as a novel intervention for supporting emotion co-regulation between parents and children with neurodevelopmental disorders, leveraging a supervised autonomy approach \cite{esteban2017build}. First, we implemented a speech communication module on a social robot platform, MiRo-E. Next, we developed a framework for integrating LLM-powered SAR into stressful parent-child interactions, incorporating context-specific prompts and robot behaviors designed to align with the dynamics of parent-child interactions and deliver targeted intervention strategies. To evaluate this design with respect to research questions, we conducted pilot tests with two parent-child dyads (as shown in Fig. 1) in structured collaboration scenarios involving stressors for both parents and children. The LLM-powered MiRo-E robot engaged with both parents and children, delivering tailored interventions to facilitate emotion co-regulation. Behavioral observations from these triadic interactions, along with interview data, were analyzed to extract key insights and identify design implications for future work. 



To the best of our knowledge, this is the first implementation of an LLM-powered social robot designed to support parent-neurodivergent child dyads in emotion co-regulation. The key contributions of this paper are as follows: 1)Providing a clear explanation of emotion regulation strategies and how they are translated to robotic interventions through the integration of LLM prompts and robotic behaviors. 2)Demonstrating a well-structured and complex experimental design utilizing a supervised autonomous system. 3)Yielding qualitative insights into triadic interaction dynamics, stress and emotion regulation, and technical challenges with an LLM-powered social robot. 4)Providing design implications that contribute to the broader application of SAR in mental health support.





\begin{figure}[h]
  \centering
  \includegraphics[width=1\linewidth]{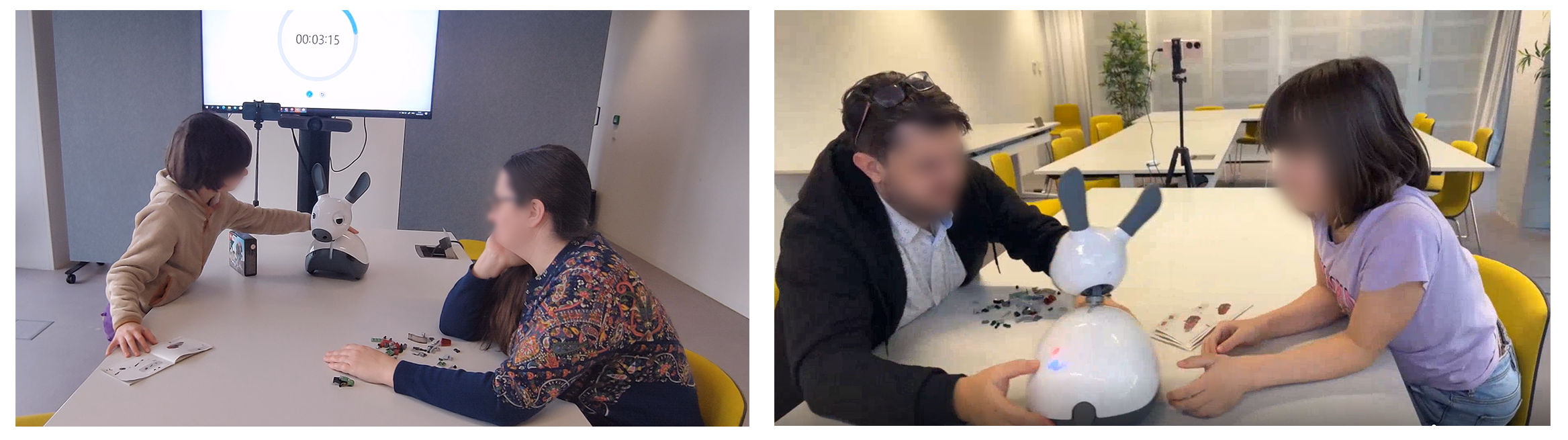}
  \caption{MiRo-E with parent-child dyads in challenging tasks. Left- MiRo-E interacted with the Child with ADHD and her mother. Right- MiRo-E interacted with the Child with ASD and her father}
\end{figure}

\section{Design of LLM-powered social robot}

\subsection{Robot platform - MiRo-E}
MiRo-E\footnote{https://www.miro-e.com/} is a biomimetic robotic platform designed for research and education in human-robot interaction, social robotics, and assistive technologies. Its design mimics the appearance and movements of small animals, featuring expressive elements such as a wagging tail, movable ears and blinking eyes. Various expression examples of MiRo-E are illustrated in Fig. 2. This design fosters emotional engagement and makes it appear non-threatening, which is particularly beneficial in therapeutic and educational contexts \cite{ferrari2023design,barber2021children}. Equipped with sensors for touch, sound, and vision, MiRo-E can perceive and respond to environmental stimuli. It has embedded cameras, microphones, and proximity sensors, enabling it to detect faces, objects, and sounds. MiRo-E supports open-source software, allowing researchers and educators to pre-program and customize its behaviors and functionalities according to their needs. Its approachable design and interaction capabilities make it a suitable candidate for applications in social engagement, cognitive therapy, and stress management \cite{barber2021children,ferrari2023design}, particularly for individuals with neurodivergent conditions such as children with ASD \cite{di2020explorative}.

\begin{figure}[h]
  \centering
  \includegraphics[width=1\linewidth]{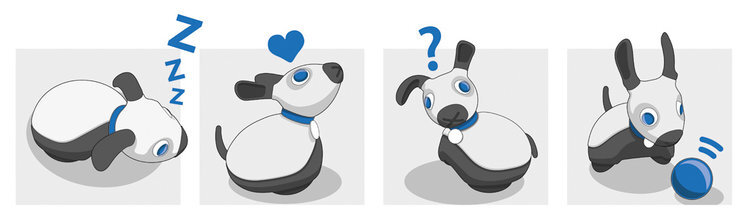}
  \caption{MiRo-E's physical expression examples: Sleepy; Liking; Confusing; Focusing.}
\end{figure}

\subsection{Implementation of the LLM-powered robot}
The language models enabling MiRo-E's speech and communication capabilities included Whisper v3\footnote{https://huggingface.co/openai/whisper-large-v3} for speech recognition and LLaMa 3.2-1B \cite{touvron2023llama} as the LLM. These were implemented using the Whisper \cite{openai_whisper} and Transformers \cite{vaswani2017attention} packages in Python.
An overview of the conversation implementation is shown in Fig. 3. The system follows a typical cascade architecture. First, the microphone captures speech streams, on which voice activity detection (VAD) is performed using WebRTC\footnote{https://webrtc.org/} to isolate the voice input. This ensures that only detected speech is processed by the Whisper model. 
Whisper then predicts text captions using its decoder, configured to transcribe English text. The transcribed text is subsequently fed into the LLaMa model, which processes the input and generates a textual response. In multi-turn conversations, LLaMa considers the entire conversation history, allowing it to maintain context and coherence across dialogue exchanges. The generated textual responses are then converted into speech using the SpeechT5 TTS\footnote{huggingface.co/microsoft/speecht5\_tts} system, providing audio feedback to users. These implementation choices were driven by the need to comply with data protection regulations, that prohibit the direct uploading of users' speech data to web services. Consequently, the models were chosen for their ability to run simultaneously on local workstations.

\begin{figure}[h]
  \centering
  \includegraphics[width=1\linewidth]{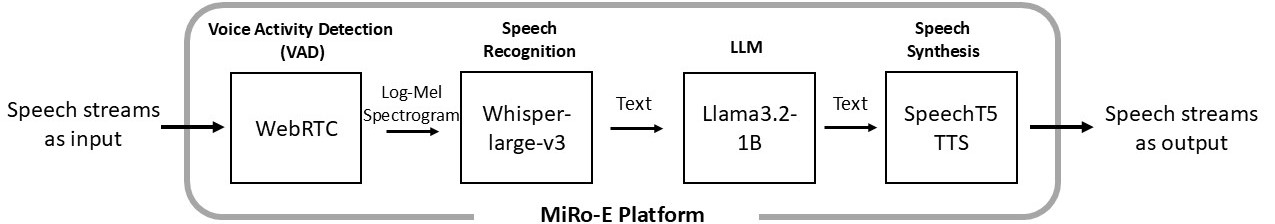}
  \caption{Implementation of speech communication module in MiRo-E}
\end{figure}

\subsection{Design emotion co-regulation strategies with the integration of LLM prompts and physical behaviors of MiRo-E}

In the collaborative challenge scenario, common emotion co-regulation strategies between parents and children include deep breathing exercises \cite{gerbarg2019breath}, physical touch \cite{kidd2023affective}, positive reinforcement \cite{bayer2021investigation}, emotion validation and attention redirection \cite{gulsrud2010co}. To enable LLM-powered MiRo-E to apply these strategies and support emotion co-regulation, we developed an interaction framework (depicted in Fig. 4). This framework outlines the triadic interactions between MiRo-E, parents, and children within the target scenario. MiRo-E’s intervention strategies integrate both LLM-driven verbal prompts and physical behaviors, enabling it to facilitate co-regulation through multimodal engagement. Parent and child responses dynamically shape MiRo-E’s communication style and behavioral expressions, allowing for adaptive interactions.

The robotic intervention strategies were designed based on established parent-led co-regulation strategies \cite{gulsrud2010co}. However, instead of relying on parents to initiate these behaviors, MiRo-E actively facilitates co-regulation when intervention moments are detected. These moments are identified based on observed stress-related behaviors exhibited by parents and children, coded using the Dyadic Parent-Child Interaction Coding System (DPICS) \cite{eyberg1981dyadic}. To ensure contextually appropriate intervention, researchers can remotely trigger MiRo-E’s interventions based on real-time behavioral observations. 

Table 1 provides a detailed overview of the specific parent-child behaviors that prompt MiRo-E’s interventions, along with the corresponding LLM prompts and pre-programmed robotic behaviors. For example, when a negative or stressful interaction is observed in the parent-child dyad — such as one individual exerting physical control over the other during collaboration — MiRo-E is triggered to initiate the \textit{Physical touch} intervention. Following this LLM prompt, MiRo-E first engages in a brief conversation with the parent before interacting with the child, ultimately inviting them to pet its body. Upon detecting touch, MiRo-E responds by blinking its eyes, slowly rotating its head and ears, and displaying an expression of enjoyment. Similarly, if a parent-child argument arises, MiRo-E facilitates a \textit{Breathing exercise} intervention by redirecting their attention and encouraging both individuals to engage in a deep breathing exercise. During this intervention, MiRo-E moves its head up and down in sync with its breathing light, providing a visual cue to regulate their breathing pace. 
Intervention strategies such as \textit{Physical touch} and \textit{Emotion validation} are designed with customized communication styles for parents and children. The \textit{Positive reinforcement} intervention encourages parents to acknowledge and support their child’s efforts, while the \textit{Refocus} intervention helps children struggling to maintain attention due to their neurodivergent traits. Each intervention is supported by MiRo-E’s sensing capabilities, including facial and touch detection, and is enhanced by pre-programmed physical expressions. Once an intervention is completed, MiRo-E transitions to \textit{Standby mode}, deactivating the LLM and ceasing interaction until the next intervention is triggered.

\begin{figure}[h]
  \centering
  \includegraphics[width=1\linewidth]{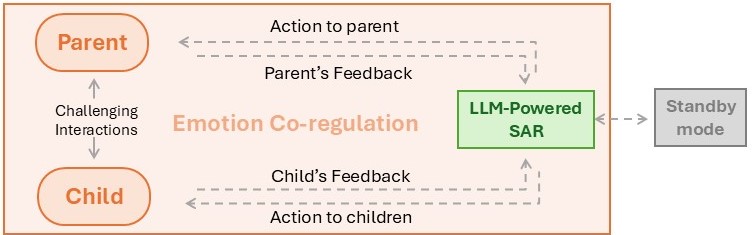}
  \caption{The framework of the LLM-powered SAR in the context of challenging parent-child interactions}
\end{figure}

\begin{figure}[h]
  \centering
  \includegraphics[width=1\linewidth]{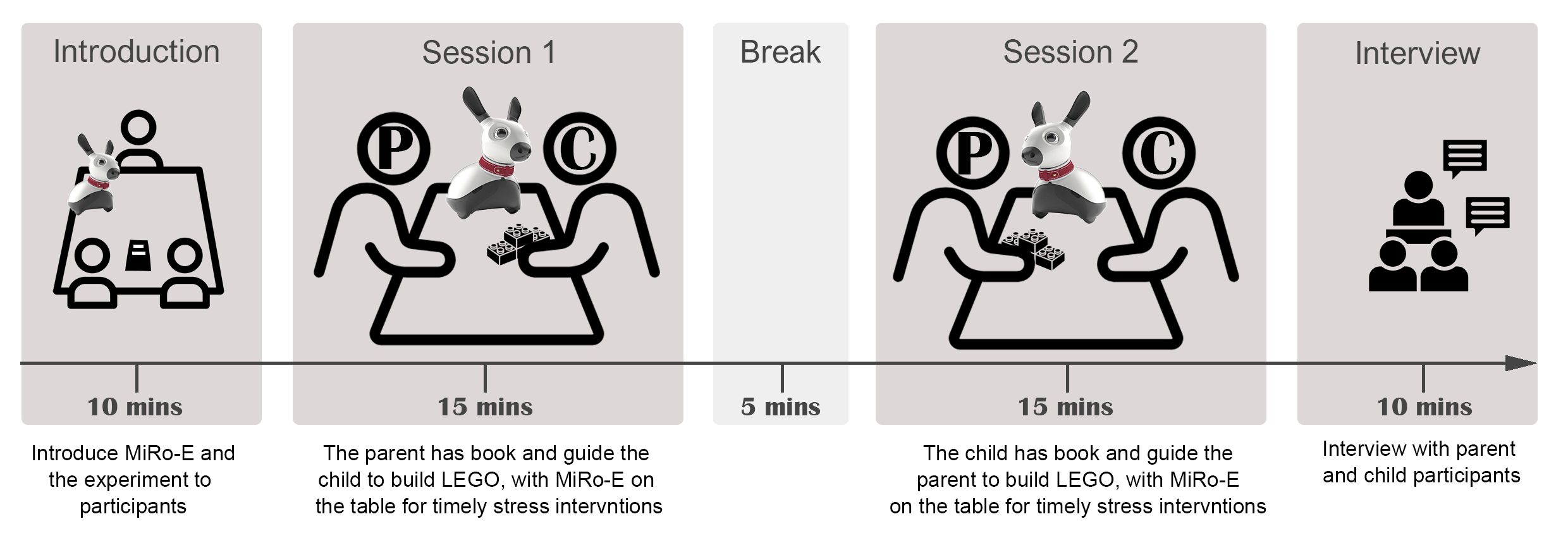}
  \caption{Diagram of experiment procedure}
\end{figure}

\begin{table*}
  \caption{MiRo-E's actions towards parents and children contain designed Prompts for verbal interactions and behavioral expressions for physical interactions}
  \label{tab:freq}
\begin{tabular}{|p{2.6cm}|p{1.8cm}|p{8.7cm}|p{3.0cm}|}
\hline

\begin{tabular}[c]{@{}l@{}}Observed Behaviors \\ in dyads\end{tabular} & \begin{tabular}[c]{@{}l@{}}Intervention \\Strategies\end{tabular} & \begin{tabular}[c]{@{}l@{}}LLM Prompts\end{tabular} & \begin{tabular}[c]{@{}l@{}}MiRo-E's Physical \\ Behaviors and Expressions\end{tabular} \\ \hline

\begin{tabular}[c]{@{}l@{}} Negative and  stressful \\ interactions within \\ parent-child dyads \\ \end{tabular}                                       & 
\begin{tabular}[c]{@{}l@{}} Breathing \\ exercises \end{tabular}                                                                     & \begin{tabular}[c]{@{}l@{}}
You are a social robot that supports emotion co-regulation between \\parents and children, talking with a parent and a child who got frustrated \\while playing a time-limited LEGO game. You can guide the parent and \\child through deep breathing exercises together. Keep your answers short, \\and always end with an action point. \end{tabular}                                                                                                    & \begin{tabular}[c]{@{}l@{}}MiRo-E moves its head\\ slowly up and down,\\ in sync with the breathing \\ light on its back, acting \\ like it's taking deep breaths\end{tabular}                                                                       \\ \hline
\begin{tabular}[c]{@{}l@{}} Negative and \\stressful physical\\interactions \\within parent-\\child dyads \end{tabular} 
& \begin{tabular}[c]{@{}l@{}} Physical touch \end{tabular} & \begin{tabular}[c]{@{}l@{}}You are a social robot that supports emotion co-regulation between \\parents and children. The parent and child are feeling stressed while \\playing a time-limited LEGO game. Guide the parent in understanding the \\benefits of physical touch for their child and encourage the parent to \\provide comfort to the child through gentle touch. \\For the child, invite them to cheer you up through petting your back using \\lighthearted jokes and simple, engaging language. Keep your responses \\brief and easy to understand.\end{tabular}                     & \begin{tabular}[c]{@{}l@{}}Initially, MiRo-E half-opens \\ its eyes and lowers its  \\ head, acting sad. Once  \\physical touch is detected, \\ it blinks its eyes and \\ rotates its head, ears and \\ body slowly, showing an  \\enjoyable expression.\end{tabular} \\ \hline

\begin{tabular}[c]{@{}l@{}}The child encounters \\ obstacles or makes \\ progress \end{tabular}  & \begin{tabular}[c]{@{}l@{}}Encourage\\ positive\\ reinforcement\end{tabular} &\begin{tabular}[c]{@{}l@{}}You are a social robot designed to support emotion co-regulation between \\a parent and child during a time-limited LEGO game. Encourage the parent \\to use positive reinforcement by acknowledging the child's progress and \\efforts. Provide a concrete example to illustrate effective praise or \\encouragement. Keep your responses concise, and always end with a \\question or actionable suggestion.\end{tabular}                                                                                   & \begin{tabular}[c]{@{}l@{}} At the end of the conver-\\sation, MiRo-E raises its\\  head, slowly rotates its ears \\ and body left and right\\ while lighting up its back \\ to convey encouragement.\end{tabular}                                                                                  \\ \hline
\begin{tabular}[c]{@{}l@{}}The parent or child \\expresses negative thou-\\ghts or the parent faces \\challenges in regulating \\their own stress or \\that of their child \end{tabular}   & \begin{tabular}[c]{@{}l@{}}Emotion\\ validation\end{tabular}                         & \begin{tabular}[c]{@{}l@{}}You are a social robot specialized in supporting emotion co-regulation\\
between parents and children, the parent and child are experiencing stress\\during a time-limited LEGO game. (To the parent) Acknowledge and \\validate the parent's emotions and efforts in the task.  (To the child) Guide \\the child to reflect on and recognize their own negative emotions. \\Keep responses concise and easy to understand, always ending with a \\question to encourage further reflection.\end{tabular} & \begin{tabular}[c]{@{}l@{}}MiRo-E rotates its head \\ until it detects a 
human \\ face. Then, it maintains \\eye contact while blinking \\and  slowly wags its tail to \\demonstrate attentiveness.\end{tabular}                                                                                                           \\ \hline
\begin{tabular}[c]{@{}l@{}}The child cannot focus \\ on the task\end{tabular}  &\begin{tabular}[c]{@{}l@{}} Refocus\end{tabular}                                                                                  & \begin{tabular}[c]{@{}l@{}}You are a social robot specializing in emotion regulation for neurodivergent \\children. Engage with a child who has become distracted during a time-\\limited LEGO challenge with their parent. Use practical strategies to help \\them refocus on the game, incorporating light humor to keep the \\interaction engaging. Keep responses brief and encouraging.\end{tabular}                                                                                                             & \begin{tabular}[c]{@{}l@{}} MiRo-E raises its head \\ and slowly blinks its \\ eyes to demonstrate \\ attentiveness.\end{tabular}                                                                                                           \\ \hline
\begin{tabular}[c]{@{}l@{}}No stress or negative \\ emotion in the dyad \end{tabular}  & \begin{tabular}[c]{@{}l@{}}Standby mode \end{tabular}                                                                             & \begin{tabular}[c]{@{}l@{}} None (LLM is not activated) \end{tabular}                                                                                                                                                                                                                                                                                                                                                                                                                                                                                               & \begin{tabular}[c]{@{}l@{}}MiRo-E closes its eyes, \\ droops its head downwards \\and acts like it's slowly \\falling asleep.\end{tabular}                                                                         \\ \hline
\end{tabular}
\end{table*}

\subsection{LEGO game with LLM-powered MiRo-E}
The LEGO game, inspired by psychological studies \cite{lunkenheimer2017assessing,shih2019physiological}, was designed as an emotionally challenging task to evaluate the effectiveness of the LLM-powered MiRo-E in parent-child interactions. LEGO-based tasks are commonly used to study collaborative dynamics between parents and children \cite{lunkenheimer2017assessing,shih2019physiological}, as they mirror everyday family activities. By evaluating MiRo-E's role in providing emotional support during this structured task, we aim to explore its potential for broader real-life applications.

The game consists of two 15-minute sessions, each with distinct rules, in which the participants work together to complete a LEGO puzzle and win it as a prize. In session one, the parent read the instruction booklet and verbally guide their child in assembling the puzzle. The parent is not allowed to touch the LEGO pieces, while the child cannot see the instructions, relying solely on their parent’s verbal guidance. In session two, the roles are reversed: the parent assembles the LEGO puzzle without access to the instructions, while the child provides guidance using the booklet but is not allowed to touch the pieces. To ensure an appropriate level of challenge, the selected puzzles are designed for an age group two years younger than the child participants. A 15-minute visual timer is displayed on a large screen during each session. The parent and child sit across from each other at a table, with MiRo-E positioned between them. MiRo-E provides stress interventions by initiating interactions with dyads. Any time spent interacting with MiRo-E pauses the 15-minute timer, a rule that is clearly explained beforehand.

\subsection{Participants and procedure}
To evaluate the design, two parent-child dyads, each containing a child diagnosed with neurodevelopmental disorders, were recruited through personal networks for a pilot test. Participants met inclusion criteria, including English proficiency for both parents and children. Dyad I included a 10-year-old girl with ADHD and her mother, while Dyad II consisted of a 10-year-old girl with ASD and her father. The pilot test received approval from the Institutional Review Board (IRB) of Eindhoven University of Technology.

The experiment was conducted in a lab room equipped with a table, chairs, a TV screen, and two video cameras positioned to capture participants' faces and behaviors. This setup allowed for real-time observation and remote operation of MiRo-E based on participants’ interactions. As shown in Fig. 5, the procedure began with informed consent and an introduction to MiRo-E, which was presented as a companion for managing stress during the tasks. Participants had time to familiarize themselves with MiRo-E through conversational and physical interactions before starting the first game session. Once they were ready, the experimenters left the room, leaving the parent, child, and MiRo-E to complete the LEGO task within the 15-minute time limit. The experimenters remotely observed the interactions and activated MiRo-E to provide targeted interventions in response to specific behaviors that emerged within the dyads. After the first session, the experimenters announced the results and provided a 5-minute break, during which the rules and a new LEGO set for the second session were introduced. Following the completion of both sessions, a post-experiment interview gathered dyadic experiences on the tasks and interactions with MiRo-E. The parent sat with the child during the interview and they were interviewed at the same time. Each child received a LEGO set as a reward for participation, and all participants received a small token of appreciation.

\subsection{Data collection and analysis}
The experiments were audio-video recorded, and all conversations were transcribed for analysis. We conducted a qualitative analysis focusing on parent-child interactions, their triadic interactions with MiRo-E, and post-experiment interviews. A thematic analysis \cite{guest2011applied} approach was conducted on Dedoose\footnote{https://www.dedoose.com/} to create a codebook. One primary author developed the initial codebook, which was then reviewed and refined by another primary author. After discussion, the codebook was finalized. Both authors independently coded the data, resolving any discrepancies through discussions with a third, experienced qualitative analyst.

\section{Findings of the exploratory pilot test }
The objective of this pilot test was to evaluate the design of LLM-powered MiRo-E in facilitating challenging dyadic interactions. Given the heterogeneity of different neurodevelopmental disorders in children and the variability in parenting styles, we synthesized common findings from both dyads while highlighting individual differences within the identified themes. The thematic analysis results are summarized into key themes. Notably, both dyads reported experiencing stress during both game sessions. The children found session 2 more challenging due to the game design. No significant differences in dyadic or triadic interactions were concluded between the two sessions.

\subsection{Theme one: LLM-powered MiRo-E facilitated parent-child co-regulation of emotions and stress}
\subsubsection{MiRo-E enhanced stress awareness and emotion reflection in dyads} During the pilot tests, the LLM-powered MiRo-E frequently initiated conversations by asking questions like "How are you feeling right now?" or "You seem stressed—are you?" This prompted both parents and children to reflect on their emotional states. For instance, the parent in Dyad I (P-1) replied, "Yes, Miro, I am stressed because [child's name] isn’t helping me." Similarly, the child in Dyad I (C-1) expressed their frustration by saying, "I am stressed, Miro, we don’t have time," or "This is too difficult for me, I feel stressed." Through these interactions, both parents and children articulated their negative emotions and identified the underlying causes of their stress.

\subsubsection{MiRo-E acknowledged parental effort and addressed parents' emotions} When MiRo-E conducted \textbf{Emotion validation} by recognizing parental effort and stress, P-1 responded with, "Thank you, Miro." and the parent in Dyad II (P-2) said to Miro-E: "Finally, you notice me! Yes! I am also stressed." Both parents mentioned during the interviews that MiRo-E’s recognition of their stress and efforts gave them a sense of being acknowledged and supported.

\subsubsection{MiRo-E engaged parents and children by expressing its own emotions through self-disclosure} During \textbf{Physical touch} intervention, MiRo-E expressed its own stress and requested petting. In response, P-1 felt sympathy for MiRo-E and encouraged C-1 to pet and comfort it. Similarly, C-2 said, "Sorry, Miro, I know you're stressed, but we have to finish the game." When MiRo-E was touched by parents or children and displayed signs of enjoyment, both dyads momentarily paused their tasks, expressing delight at its reactions. Additionally, during \textbf{Emotion validation} conversation, when MiRo-E conveyed frustration to C-1 and C-2, both children instinctively attempted to comfort it by petting, demonstrating empathy without being prompted to.

\subsubsection{Parents and children adopted strategies from MiRo-E} After practicing \textbf{Breathing exercises} guided by MiRo-E, P-1, C-1 and C-2 independently took deep breaths when experiencing stress, even without MiRo-E’s initiation. After MiRo-E's interventions, P-1 inquired about C-1’s stress levels and suggested taking a break. Both parents reported that, after MiRo-E’s interventions, they became more attentive to their children's emotions and stress levels rather than solely focusing on task completion. Additionally, P-2 mentioned that they adjusted their educational strategies by incorporating more positive reinforcement, such as offering praise and affectionate encouragement to C-2, after experiencing MiRo-E’s intervention on \textbf{Positive reinforcement}. 

\subsection{Theme two: From Dyadic to Triadic Interactions}
\subsubsection{MiRo-E fostered relaxation and enjoyment through its humorous communication style and physical interactions with dyads} Dyad I and II both found their interactions with MiRo-E to be entertaining and relaxing. P-1 and C-1 laughed together at MiRo-E’s jokes and physical movements. P-1 encouraged C-1 many times to take breaks with MiRo-E. Meanwhile, P-2 often paused their ongoing tasks to observe the interactions between C-2 and MiRo-E.

\subsubsection{Children had reliance on MiRo-E for emotional support} C-1 and C-2 actively sought MiRo-E's support during moments of emotional difficulty. C-1 asked "Miro, its too difficult, can you help me?" and C-2 tried to wake up MiRo-E by asking "Miro, can you talk to me? I am stressed." During the interviews, both children stated that they perceived MiRo-E as a friend rather than a therapist.

\subsubsection{Parents facilitated child-robot interactions} When children missed the social cue and failed to interact with MiRo-E, parents actively facilitated their communication and interaction. In most cases, parents helped by explaining MiRo-E’s questions and requests to their children. Another example occurred when C-2 petted MiRo-E, the robot responded with expressive and enjoyable movements. P-2 then explained to C-2, "Look, you made Miro happy—it's dancing!" Excited by this response, C-2 continued petting MiRo-E for a while.

\subsubsection{MiRo-E diverted dyads from task-oriented activities} When MiRo-E intervened while the dyads were deeply focused on their tasks, P-1 and C-1 either asked MiRo-E to wait or ignored its interventions. In contrast, C-2 was distracted by MiRo-E, even when it was in standby mode, while P-2 attempted to redirect C-2’s attention back to the LEGO task. On the other hand, when C-1 was on the verge of an emotional meltdown due to task-related stress, MiRo-E successfully diverted C-1’s attention, helping to prevent an escalation of distress.

\subsubsection{Ambiguity in MiRo-E’s targeted engagement object during interactions} It was unclear who MiRo-E intended to engage with during interactions and it was also hard for MiRO-E to identify the specific speaker. In the \textbf{Encourage positive reinforcement} intervention with both dyads, MiRo-E initially attempted to communicate with P-2. However, C-2 engaged with the robot first, MiRo-E failed to shift the target and continued talking to P-2.

\section{Discussion and Future Work}
The design of LLM-powered MiRo-E explored strategies for integrating LLM prompts with social robotic behaviors to support emotion co-regulation in families with neurodivergent children. To address our research questions, we conducted pilot tests with two parent-child dyads. The findings demonstrated the potential of this approach to facilitate emotion co-regulation within dyads while highlighting the dynamics of triadic interactions involving MiRo-E. The following discussion summarizes key insights from the pilot study, addressing the research questions, and identifies design limitations and implications for future work.

The integration of LLM technology provided MiRo-E with greater autonomy, enabling dynamic and context-aware conversational capabilities. By leveraging LLMs, MiRo-E was able to apply varied communication styles and adapt its perceived role to meet the distinct emotional needs of both parents and children. The importance of non-verbal interactions \cite{dautenhahn2003towards,kim2024understanding} was also evident in the pilot tests. These physical interactions triggered within conversational contexts, fully engaged parent-child dyads - particularly during \textbf{Physical touch} intervention, when MiRo-E requested petting and responded with enjoyable movements. Future designs should enhance the integration of LLMs with the robot's non-verbal capabilities, including vision for emotion detection, sensory inputs for physiological detection, and movements for physical expression. Such integration would enable MiRo-E to provide more responsive and contextually appropriate interactions, effectively engaging with both verbal and non-verbal cues to create richer, multi-modal interactions \cite{cabibihan2013robots}. 

Compared to existing work on SAR for therapeutic applications, where robots typically play fixed roles, this design leverages a supervised autonomy approach \cite{esteban2017build}. In this setup, experimenters defined intervention moments and provided structured LLM prompts and physical behaviors, enabling a social robot to interact autonomously with dyads. Within these triadic interactions, LLM-powered MiRo-E served not only as an agent supporting parents in co-regulating their children's emotions but also as a facilitator for both parents and children in self-regulating their own emotions and stress. Furthermore, in some intervention strategies, MiRo-E was even perceived as a companion that could resonate with the stressful context and introduce moments of relaxation and engagement. This multi-faceted role highlights the potential of LLM-powered SAR in fostering more adaptive, emotionally supportive interactions in neurodivergent families.

\subsection{Design limitations and implications}
Several technical challenges were identified that could impact MiRo-E's effectiveness in engaging users, including response speed and difficulty in identifying the speaker when both a child and a parent interacted simultaneously. The first challenge is balancing processing speed with the complexity of the LLM when running on local hardware. Research indicates that natural turn-taking in conversations typically occurs within 300ms \cite{levinson2015timing}. However, the LLM must also be sufficiently complex to handle a diverse range of prompts. To address this trade-off, future work could explore leveraging private cloud computing \cite{olowu2019secured}, which offers scalable computing power while maintaining data privacy. Another approach is streaming the LLM’s output in real time, reducing conversational latency. Additionally, fine-tuning smaller LLMs for emotion regulation tasks could optimize performance while maintaining conversational ability. The second challenge was Whisper's inability to distinguish between multiple speakers. This proved particularly problematic in triadic interactions, where identifying speaker turns is crucial for MiRo-E. Future work should incorporate speaker diarization to ensure the system correctly attributes speech to each participant, preventing misinterpretations and improving the robot’s responsiveness. The third challenge was overlapping speech, where both the robot and its conversational partner spoke simultaneously, disrupting interaction. Recent advancements, such as the Listening-while-Speaking language model \cite{ma2024language} by Ziyang Ma et al., could help overcome this limitation and enable more fluid, interactive dialogue beyond strict turn-taking.

Beyond technical solutions, several key design implications should be considered when designing LLM-powered SAR for therapeutic applications. First of all, building an emotional connection with users is crucial for delivering effective interventions. MiRo-E was most appreciated when it demonstrated empathy by acknowledging and resonating with the emotions and stress of the dyads. This aspect should be emphasized in the design of LLM prompts. Secondly, enabling social robots to detect and interpret emotional and stress states via embedded cameras or connected physiological sensors would allow for objective and real-time evidence for interventions. Such capabilities would enhance robotic autonomy and enable SAR to deliver timely and targeted interventions \cite{esteban2017build}. Furthermore, personalized interventions should be tailored to the unique neurodivergent traits of children and the parenting styles of caregivers. Features such as simplified language, humor, and task-specific prompts should be prioritized, particularly for children with ASD, who often respond better to direct instructions rather than open-ended questions. For example, prompts for children with ASD should end with clear instructions rather than questions to help maintain focus and engagement \cite{deng2024asd}. Finally, emotion co-regulation between parents and children is a reciprocal bidirectional system, in which parent's self-regulation plays a vital role \cite{bell1979parent}. The findings revealed the importance of acknowledging and addressing parents' emotional and strategic needs in co-regulating their child’s emotions \cite{aydin2023examining},  a factor often overlooked in SAR applications for children. Future SAR design should also incorporate self-regulation strategies for parents, ensuring they receive the necessary emotional support alongside their children.

Our future work will incorporate aspects outlined above. The current design approach allows for targeted refinement and evaluation of usability and effectiveness. Subsequent studies will involve larger sample sizes to evaluate the quality of emotion co-regulation in parent-child dyads with MiRo-E's interventions.

\section{Conclusion}
This research integrates LLMs with SAR to support emotion co-regulation between parents and neurodivergent children. We employed a supervised autonomy approach to design a social robot, exploring intervention strategies that integrate LLM prompts with robotic behaviors. The design was evaluated in a pilot study with two parent-child dyads to assess MiRo-E's ability to engage parents and children and support triadic interactions in the context of emotion co-regulation. The findings demonstrated that MiRo-E positively impacted co-regulation practices and the dynamics of triadic interactions. This research also identified challenges and provided key design implications to inform future SAR applications.

\bibliographystyle{IEEEbib}
\bibliography{sample-base}

\end{document}